\documentclass[10pt,journal]{IEEEtran}
\usepackage{cite}
\usepackage{graphicx}
\usepackage{psfrag}
\usepackage{subfigure}
\usepackage{url}
\usepackage[cmex10]{amsmath}
\usepackage{array}
\usepackage{graphics}
\usepackage{epsfig}
\usepackage{amsbsy}
\usepackage{amssymb}
\usepackage{amsthm}

\usepackage{color}
\usepackage[hidelinks]{hyperref}

\IEEEoverridecommandlockouts
\begin{document}
\title{\LARGE Energy Harvesting Cognitive Radio with Channel-Aware Sensing Strategy}

\author{Jeya Pradha~J,
        Sanket S.~Kalamkar,~\IEEEmembership{Student~Member,~IEEE,}
        and~Adrish~Banerjee,~\IEEEmembership{Senior~Member,~IEEE}
        \vspace*{-4mm}

\thanks{The authors are with EE department, IIT Kanpur, 208 016, India. (e-mail:  jeyapradhaj@gmail.com, $\lbrace$kalamkar, adrish$\rbrace$@iitk.ac.in).}
}

\maketitle

\begin{abstract}
\boldmath
An energy harvesting cognitive radio scenario is considered where a secondary user (SU) with finite battery capacity opportunistically accesses the primary user (PU) channels. The objective is to maximize the throughput of SU under energy neutrality constraint and fading channel conditions in a single-user multi-channel setting. Channel selection criterion based on the probabilistic availability of energy with SU, channel conditions, and primary network's belief state is proposed, which chooses the best subset of channels for sensing, yielding higher throughput. We construct channel-aware optimal and myopic sensing strategies in a Partially Observable Markov Decision Process framework based on the proposed channel selection criterion. The effects of sensing errors and collisions between PU and SU on the throughput of latter are studied. It is shown that there exists a trade-off between the transmission duration and the energy lost in collisions.
\end{abstract} 

\vspace*{-4mm}

\section{Introduction}
\PARstart{C}{ognitive} radio (CR) with energy harvesting (EH) capability is a way to overcome spectrum scarcity problem while achieving green communications \cite{park1}. In CR, when the spectrum access for a secondary user (SU) is opportunistic and restrained by sensing errors, the possibility of transmission, in turn, the throughput of SU is limited. Moreover, the time-varying and random nature of fading channels may significantly reduce the SU throughput. Thus, it is imperative to access the channels with better conditions and low primary user (PU) occupancy, emphasizing that the design of channel sensing strategies should take these factors into account. 

The optimal and sub-optimal sensing strategies for CR \textit{ad hoc} networks in an unconstrained energy setting over finite horizon have been developed in \cite{zhao_adhoc}. In \cite{park}, the proposed sub-optimal policy in \cite{zhao_adhoc} is analyzed for EH opportunistic spectrum access (OSA) based CR networks under perfect sensing.  A single-user multi-channel setting is considered in \cite{zhao_adhoc, park} in a Partially Observable Markov Decision Process (POMDP) framework. In \cite{sultan_8}, spectrum sensing policies are studied for energy-constrained CR taking into account the dynamics of the primary network in a POMDP framework. Energy harvesting CR which optimizes its sensing and transmit energies is analyzed in \cite{sultan} for a single-user single-channel setting in the presence of sensing errors.

The main contributions of this letter are as follows. Firstly, under energy neutrality constraint \cite{kansal}, we propose a channel selection criterion for an energy harvesting CR to maximize the average spectral efficiency of SU. The proposed criterion exploits not only the knowledge of PU occupancy and channel conditions taking sensing errors into account, but also the dependency of the decision of SU to sense and access PU channels on the probabilistic availability of energy with SU. Secondly, based on the proposed channel selection criterion, we develop optimal and myopic policies in a POMDP framework for SU to choose the channel(s) for sensing. Thirdly, we highlight the effect of sensing errors on SU's average spectral efficiency through the trade-off between the transmission duration and the energy lost in collisions between PU and SU. \vspace*{-4mm}
 
\section{System Model}
Consider a spectrum consisting of $N$ channels each of bandwidth $B$ licensed to a primary network (PN). The occupancy of PN follows a discrete-time Markov process with $2^{N}$ states. Both PN and SU are assumed to follow time-slotted synchronous communication \cite{zhao_adhoc}. SU is embedded with energy harvesting capability and has finite battery capacity $e_{\text{max}}$. The energy harvesting process is assumed to be stationary and ergodic with mean $P_{EH}$ J/s similar to \cite{parag}. The Bernoulli model is used for illustration, which is as follows: In each time slot, SU harvests energy $e_{h} $ with probability $p_{h}$, thus, ${P}_{EH} = p_{h} e_{h}$. 

We model SU in OSA paradigm, in which the PU channels are accessed opportunistically with sufficient protection to PU by satisfying the target probability of detection $ P_{D}$ and the probability of false alarm $P_{F}$. For a given channel gain $\sqrt{h}$ between PU and SU, to satisfy the target $ P_{D}$ and $P_{F}$ when PU signal and noise are zero mean circularly symmetric complex Gaussian (CSCG), the minimum number of samples required for sensing using energy detection is \cite{rugini1}
\begin{equation}
L_{s,\min} = \left\lceil (1/36)\left(p + \sqrt{p^{2} + 4}\right)^{2}\right\rceil,
\label{eqn:nmin}
\end{equation}
where {{\small{$p = ((1 + \gamma \vert h \vert)^{-1/3}Q^{-1}(P_{F}) - Q^{-1}(P_{D}))/(1 - (1 + \gamma \vert h \vert)^{-1/3})$, $ \gamma$}}} denotes the received signal-to-noise ratio of PU signal at SU and {{\small$Q(x) = (1/\sqrt{2\pi}) \int_{x}^{\infty}\exp( -u^{2}/{2}) du$}}. Then the minimum sensing time is
\begin{equation}
T_{s, \min}  = L_{s,\min}/f_{s},
\label{eqn:tmin}
\end{equation}
where $f_{s} $ is the sampling frequency. The energy required for sensing is the product of number of samples $ L_{s} $ and the energy required for sensing each sample $ e_{s,\text{sample}} $, which is given by
\begin{equation}
e_{s} = L_{s} \times e_{s,\text{sample}}.
\label{eq:sens_ene}
\end{equation}
Once the PU channel is found idle, SU may transmit over the channel. We consider frequency-flat, block Rayleigh fading channel of bandwidth $B$ and coherence time $T_{c}$ between secondary pair. The SU transmitter sends pilot of duration $T_{est}$ to the SU receiver and gets the perfect channel state information (CSI) through an error-free and dedicated feedback channel. The received signal $v$ at the SU receiver in a time slot is $v = \sqrt{g}u + n$
where $g$ is the channel power gain, $u$ is SU's transmitted signal and $n$ is CSCG noise with power spectral density $N_0/2$. We consider the case that SU always has data to transmit in order to determine the maximum achievable throughput. The receiver acknowledges the transmitter by an error-free ACK (ACK = 1 for successful transmission) on a dedicated feedback channel at the end of the slot\cite{sultan}. The ACK is assumed of very short duration compared to slot duration $ T $. An unsuccessful transmission occurs only when SU's transmission collides with PU's transmission.

The EH secondary transmitter adapts the transmit power and rate according to the channel power gain $g$ and the available energy $e$. SU adopts a $M$-QAM constellation set $ M \in \lbrace  M_{1}, \dotsc, M_{K} \rbrace $ for some $ K $, where $M_k = 2^{2(k - 1)}$, $k = 1, \dotsc, K$ and $M_{1}$ corresponds to no transmission. The channel power gain levels are divided into $ K $ regions denoted by $\mathcal{G}_{k}$ and each region is mapped to a constellation size $M_{k}$ \cite{parag}. Then the throughput is $\log_2M_k$. 
The transmit power $P_{tr}$ for given $M_k$ and $g$ at time slot $t$ is\cite{parag} 
\begin{equation}
P_{tr}(g, M_{k}) =  \frac{-\ln(P_{b}/c_{1})}{c_{2}}(2^{c_{3}\log_{2}M_{k}}-c_{4})\frac{N_{0}B}{g},
\label{eq:ptr}
\end{equation}
where $P_{b}$ is the bit error rate, $c_{1}=$ 2, $c_{2}=$ 1.5, $c_{3}=$ 1, $c_{4}=$ 1.  
The energies required for transmission $e_{tr}$ and for circuit operation $e_{ckt}$ are given by 
\begin{equation}
e_{tr} = P_{tr}T_{tr} \hspace{2mm} \text{and}\hspace{2mm} e_{ckt} = (P_{ckt}+\kappa P_{tr})T_{tr},
\label{eq:energ}
\end{equation}
where $1/(1 +\kappa)$ is the power drain efficiency of power amplifier \cite{energy_goldsmith}, $T_{tr}$ is the transmission duration in a time slot, $P_{ckt}$ is the power required for circuit operation and $\kappa P_{tr} $ denotes the power consumed by the power amplifier.\vspace*{-4mm}

\section{Problem Formulation}
We assume that SU has partial knowledge about PN as SU may not be able to sense all the channels in PN due to hardware and energy constraints. Also, sensing errors add uncertainty about PN. The PN occupancy is partially observable, whereas the residual energy and channel power gain are fully observable. Note that $g$ is estimated, i.e., fully known only if $e \geq e_{est}$, where $e_{est}$ is the energy required for estimation.\vspace*{-4mm}
\subsection{System Components}
The PN occupancy during time slot $t$ is given by $ \textbf{s} = [s_{1}, s_{2},\dotsc, s_{N}] $ where $s_{i} \in \lbrace 0 \hspace{1mm} (\text{occupied}), 1 \hspace{1mm} (\text{idle}) \rbrace$. The state of the secondary network $\mathcal{S}$ is characterized by PN occupancy $\textbf{s}$, energy available in the battery $e$ and the channel power gain $g$ between the secondary pair. It can be defined as
\begin{equation}
 \mathcal{S}\triangleq \lbrace (\textbf{s},e,g) : \textbf{s} \in \lbrace 0,1\rbrace ^{N}, e \in [0, e_{\max}], g \in [0, \infty) \rbrace.
 \label{eq:srrr}
\end{equation}
Let $\Lambda_0$, $\Lambda_1$ and $\Lambda_2$ be the sets of estimated channels, channels to sense and channels to access, respectively. At the beginning of a slot, SU can either remain idle or perform the following sequence of operations: 
(i) Estimate upto $ \vert\Lambda_{0}\vert$ $(\vert\Lambda_{0}\vert \leq N )$ channels;
(ii) Sense upto $ \vert\Lambda_{1}\vert$ $(\vert\Lambda_{1}\vert \leq \vert\Lambda_{0}\vert )$ estimated channels;
(iii) Access upto $ \vert\Lambda_{2}\vert$ $(\vert\Lambda_{2}\vert \leq \vert\Lambda_{1}\vert )$ sensed channels.
Let $a \in \lbrace 0 \hspace{1mm} (\text{idle}) , 1, \dotsc, N \rbrace$, $\hat{a} \in \lbrace 0 \hspace{1mm} (\text{idle}), 1, \dotsc, N \rbrace$, $d_{\hat{a}} \in \lbrace 0 \hspace{1mm} (\text{no access}), 1 \hspace{1mm}(\text{access}) \rbrace$ denote the indices of estimated channels, channels available for sensing and access decision, respectively.
SU observes the channel either as occupied (0) or idle (1) after sensing. The probability of the sensing observation $ o$, given the channel $ \hat{a} $ is sensed with errors is 
\begin{equation}
P(o_{\hat{a}} = 0) =
  P_{D}\text{I}_{s_{\hat{a}} = 0} + P_{F}\text{I}_{s_{\hat{a}} = 1} \hspace{1.9 cm}  \text{if} \hspace{0.2 cm} \hat{a} \neq 0, 
  \label{obs_0}
\end{equation}
\begin{equation}
P(o_{\hat{a}} = 1) = 
  (1 -  P_{D})\text{I}_{s_{\hat{a}} = 0} + (1-P_{F})\text{I}_{s_{\hat{a}} = 1} \hspace{0.1 cm} \text{if} \hspace{0.2 cm} \hat{a} \neq 0,
   \label{obs_1}
\end{equation}   
where $\text{I}_{{x}}$ denotes the indicator function which takes value $1$ if $x$ is true, otherwise $0$; $ s_{\hat{a}}$ is the state of the channel $\hat{a}$.

The probability that PN occupancy transits to state $\textbf{s}^\prime$ at the beginning of time slot $t+1$ from $\textbf{s}$ at time slot $t$ is denoted by $P_{\textbf{s}^\prime,\textbf{s}}$. 
The energy available $e^\prime$ at the start of time slot $t+1$, is dependent on the energy available $e$, energy consumed $e_{c}$ and energy harvested $e_{h}$ at slot $t$. Then
\begin{equation}
P(e' \mid e ) = \left\{
  \begin{array}{l l}
    p_{h} & \quad  e^\prime = \min(e - e_{c} + e_{h}, e_{\max}),\\
    1-p_{h} & \quad   e^\prime = e - e_{c}, 
    \end{array} \right.
    \label{eq:43}
\end{equation}
with
\begin{equation}
e_{c}=\left\{
  \begin{array}{l l}
    e_{est}+e_{s}+e_{ckt}+e_{tr} & \quad \text{if} \hspace{0.2 cm}  \hat{a} \neq 0 $, $ o_{\hat{a}} = 1,\\
    e_{est}+e_{s} & \quad \text{if} \hspace{0.2 cm}    \hat{a} \neq 0 $, $ o_{\hat{a}} = 0, \\
    e_{est}  & \quad \text{if} \hspace{0.2 cm}  \hat{a} =0 ,\\
    e_{i} & \quad \text{if} \hspace{0.2 cm}   a = 0 , \\    
    \end{array} \right.
    \label{eq:ene}
\end{equation}
where $e_{est} $, $ e_{s}$, $e_{ckt} $ and $e_{tr} $ are given by \eqref{eq:est_energ}, \eqref{eq:sens_ene} and \eqref{eq:energ}, respectively. $e_i$ is the energy consumed when SU is idle and it is considered to be negligible. We assume that the channel power gain $g'$ at time slot $t+1$ is independent of the channel power gain at previous time slots, i.e., $g'$ can be traced to any of the $K$ fading regions\cite{parag}.

SU maintains a belief vector $\text{\textbf{b}}= [\textit{b}({1}), \textit{b}({2}),\dotsc, \textit{b}({2^{N}})] $ about PN, where $\textit{b}({i})$ is the conditional probability that the network state is $i$ given all the past decisions and observations.\vspace*{-4mm}

\subsection{Channel Selection Criterion}
We propose a channel selection criterion as a function of belief about PN occupancy and the energy-constrained spectral efficiency $\eta$. Given $e$ and $P_{b}$, for a channel with power gain $g$, $\eta = \log_2M_k$ if $g \in \mathcal{G}_{k}$ and $e_{s} + e_{ckt} + e_{tr} \leq e-e_{est}$; $\eta$ is zero when $g \in \mathcal{G}_{1}$ or/and $ \text{I}_{e \geq e_{c}} =0$.

At a time slot $t$, the reward $R_{\hat{a}}$ on a channel $\hat{a}$ is defined as the spectral efficiency $\eta_{\hat{a}}$ for that channel $\hat{a}$  and is given by
\begin{equation}
R_{\hat{a}}(t) = \left\{
  \begin{array}{l l}
    s_{\hat{a}} \times \eta_{\hat{a}}(t)  & \quad \text{if} \hspace{0.2 cm} \hat{a} \neq 0, \text{ACK} = 1,\\  
    0 & \quad \text{if} \hspace{0.2 cm} \hat{a} = 0 \hspace{0.2cm}\text{or ACK} \neq 1.\\
  \end{array} \right. 
  \label{reward}
  \end{equation}
The channel will be sensed if and only if $\eta$ on that particular channel is non-zero. The optimal and myopic policies to choose the channel(s) for sensing are as follows:

\subsubsection{Optimal Policy}
The value function for a finite horizon $T_{F}$ including the partially and fully observable system components can be formulated as\vspace*{-2mm}

{{\small
\begin{multline}
\hspace{-1mm}V^{*}_{t=T_{F}}(\text{\textbf{b}},e,g) = \max\limits_{\hat{a}}\sum_{\textbf{s}' \in \textbf{S}}b(\textbf{s}')\sum_{\textbf{s} \in \textbf{S}}P_{\textbf{s}',\textbf{s}}\sum_{z=0}^{1}P(o = z \mid \textbf{s}, \hat{a})\\\times \left( zR_{\hat{a}}(t) + \sum_{e'}\sum_{\mathcal{G}_{1}}^{\mathcal{G}_{K}} P(e' \mid a, \hat{a},  o_{\hat{a}}, e) P(g') V^{*}_{t-1}(\text{\textbf{b}}', e', g' )\right),
\label{eqn:optimal}
\end{multline}}}

\noindent with $\vert \textbf{S} \vert = 2^{\vert \Lambda_{0} \vert}$, $V^{*}_{t-1}$ is the maximum expected reward that can be accrued over $t-1$ remaining slots and the updated belief $b'$ obtained using Bayes' rule is as follows:

{{\small
\begin{eqnarray*}
&&\textit{b}'(\textbf{s}) \\&&= 
\left\{
  \begin{array}{l l}
    \sum_{\textbf{s}'}b(\textbf{s}')P_{\textbf{s}',\textbf{s}}, \hspace{2.1 cm}  \hat{a} = 0  \hspace{0.1 cm}\text{or} \hspace{0.1 cm} a=0, \\
   \frac{\sum_{\textbf{s}'}b(\textbf{s}')P_{\textbf{s}',\textbf{s}}I_{s_{\hat{a}}=1}}{\sum_{\textbf{s}'\in \textbf{S}}\sum_{ \textbf{s}'' \in \textbf{S}}b(\textbf{s}') P_{\textbf{s}',\textbf{s}''}I_{s''_{\hat{a}}=1}}, \hspace{1mm} \hat{a} \neq 0, o_{\hat{a}} = 1, \hspace{0.005 cm} \text{ACK} = 1,  \\
   \frac{\sum_{\textbf{s}'}b(\textbf{s}')P_{\textbf{s}',\textbf{s}}I_{s_{\hat{a}}=0 }}{\sum_{\textbf{s}'\in \textbf{S}}\sum_{ \textbf{s}'' \in \textbf{S}}b(\textbf{s}') P_{\textbf{s}',\textbf{s}''}I_{s''_{\hat{a}}=0 }}, \hspace{1mm}  \hat{a} \neq 0,  o_{\hat{a}} = 1, \hspace{0.005 cm} \text{ACK} \neq 1,  \\
    \frac{\sum_{\textbf{s}'}b(\textbf{s}')P_{\textbf{s}',\textbf{s}}I_{s_{\hat{a}}=1} P(o_{\hat{a}}=0 \mid s_{\hat{a}})}{W},\hspace{0.1 cm} \hat{a} \neq 0, \hspace{0.005 cm} o_{\hat{a}} = 0,
    \end{array} \right.
\end{eqnarray*}}}

\noindent where {{\small$W = \sum_{\textbf{s}'\in \textbf{S}}\sum_{ \textbf{s}'' \in \textbf{S}} b(\textbf{s}')P_{\textbf{s}',\textbf{s}''} (\text{I}_{s''_{\hat{a}}=1} P(o_{\hat{a}}=0 \mid s''_{\hat{a}})+\text{I}_{s''_{\hat{a}}=0} P(o_{\hat{a}}=0 \mid s''_{\hat{a}}))$}}. The optimal policy chooses the channel for sensing that maximizes the expected reward, in turn, the throughput over $T_F$ slots. In the presence of sensing errors, to obtain the optimal access policy, the collision probability $ P_{col} $ between PU and SU should be equal to $ 1- P_{D} $ \cite{zhao_joint}. Then, the access decision is same as that of sensing observation and it is represented as $d_{\hat{a}} = \text{I}_ {\lbrace o_{\hat{a}} = 1 \rbrace }$. Finding the optimal policy for a POMDP is computationally prohibitive as the complexity grows exponentially with $N$ and is $\mathcal{O}(N^{T_F})$\cite{zhao_adhoc}, \cite{djonin}. Hence, we consider a myopic policy with reduced state space whose complexity increases linearly with $N$, i.e., the complexity is $\mathcal{O}(N)$\cite{zhao_adhoc}. The detailed analysis about the complexity of the optimal policy can be found in \cite{djonin}.

\subsubsection{Myopic Policy}
In myopic policy \cite{zhao_adhoc}, neglecting the impact of current action on future slots, the sensing strategy aims to maximize the expected reward only on the current slot. For this case, the expected reward for the channel $\hat{a}$ is $\left(\pi_{\hat{a}}(t)\beta_{\hat{a}}+(1-\pi_{\hat{a}}(t))\alpha_{\hat{a}}\right) \times \eta_{\hat{a}}(t)$ where $\pi_{\hat{a}}(t)$ is the belief that the channel $\hat{a}$ is idle at time slot $t$, $\beta_{\hat{a}}$ is the probability that the channel $\hat{a}$ remains in state 1 (unoccupied) and  $\alpha_{\hat{a}} $ is the probability that the channel transits from state 0 (occupied) to state 1. The belief that PU is idle at the beginning of slot $ t+1 $ before state transition is given by Bayes' rule as
\begin{equation*}
\pi_{\hat{a}}(t+1)= \left\{  
    \begin{array}{l l}
	1, &  \hat{a}(t)\neq 0, o_{\hat{a}(t)}=1, \text{ACK}=1, \\
	0, & \hat{a}(t)\neq 0,  o_{\hat{a}(t)}=1, \text{ACK} \neq 1, \\
	\frac{XP_{F}}{XP_{F}+YP_{D}}, & \hat{a}(t)\neq 0, o_{\hat{a}(t)}=0, \\
    X,  & \hat{a}(t)= 0, 
  \end{array} \right. 
  \label{eq:myo_be}
\end{equation*}
where $ X = \pi_{\hat{a}}(t)\beta_{\hat{a}}+(1-\pi_{\hat{a}}(t))\alpha_{\hat{a}} $ and $ Y = \pi_{\hat{a}}(t) (1-\beta_{\hat{a}})+(1-\pi_{\hat{a}}(t))(1-\alpha_{\hat{a}})$. The access decision is same as that of the optimal policy. Based on the expected reward, the best channel among $|\Lambda_1|$ channels is sensed first and if found busy, only then the next best channel is sensed.\vspace*{-2mm}
\section{Simulation Results and Discussion} 
The simulation parameters are assumed without loss of generality (see Table \ref{tab:simparam}). The power required for estimation $P_{est}$ is 20\% of the power $\overline{P}$ required to transmit average constellation size for an average channel power gain. Then, the energy required to estimate a channel is 
\begin{equation}
e_{est} = P_{est}T_{est},
\label{eq:est_energ}
\end{equation}
where $T_{est} = 14 T_{sym}$ as $14$ pilot symbols are sent for estimation per slot per channel and $B = 1/{T_{sym}}$. SU may sense multiple channels, but may use the best available channel for its transmission due to energy constraint, i.e., $\vert \Lambda_{2}\vert= 1 $. We take $P_{col}$ = $1 - P_D$, $T_{F} = 5$ and $e_{\text{max}} = 10e_{h} $. The number of iterations for Monte Carlo simulation is $ 10^{5} $.

\begin{figure}
\centering \hspace*{-15mm} 
      \subfigure[Comparison of optimal and myopic policies.
]{\label{diff_Tsen}\includegraphics[scale=0.31]{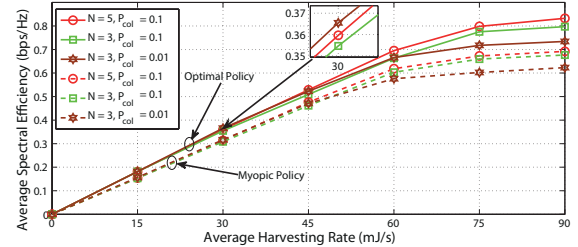}}
     \hspace*{-15mm} \subfigure[Effect of $P_{col}$ and $P_{EH}$ for myopic policy, $N$ = 5.]{\label{collvspeh}\includegraphics[scale=0.31]{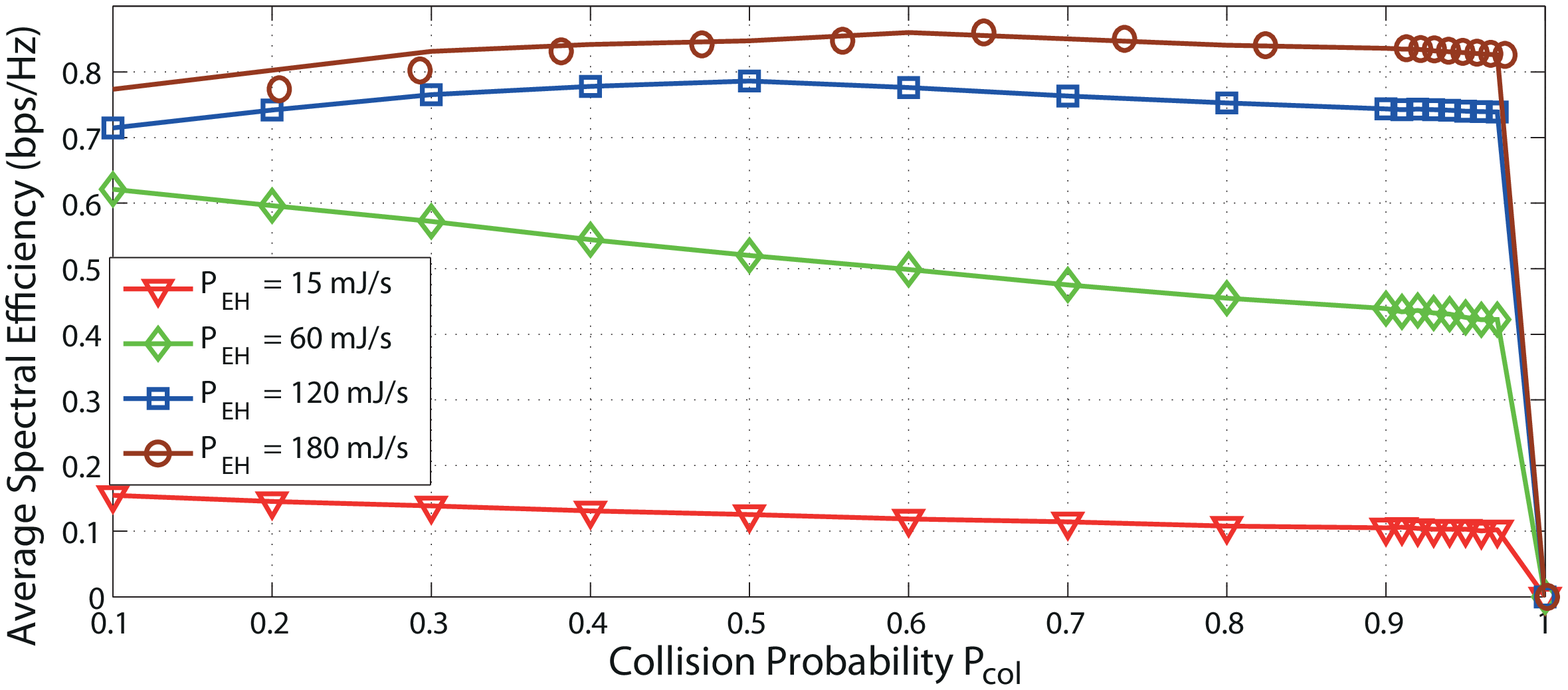}}
      \caption{Effect of $P_{col}$ and harvesting rate $P_{EH}$ on SU throughput when the channel between the secondary pair is Rayleigh with power gain $g$ and the channel between PU and SU is AWGN $\left(\sqrt{h} = 1\right)$, $P_{F} = 0.1$, $N$ = $|\Lambda_0|$, $\vert\Lambda_1 \vert$ = 1. Each channel has transition probabilities $\beta = 0.7$ and $\alpha = 0.5$.}
      \label{fig:ORrule}\vspace*{-4mm}
    \end{figure}

Fig. \ref{diff_Tsen} compares the optimal and myopic policies. It can be seen that with increase in $N$, the throughput (average spectral efficiency) increases as SU has more number of PU channels to consider for sensing increasing the probability of choosing a channel with better gain. In Figs. \ref{diff_Tsen} and \ref{collvspeh}, we have shown how the throughput varies with the collision probability $P_{col}$ for different harvesting rates $P_{EH}$, when the sensing duration is kept to its required minimum given by \eqref{eqn:tmin}. At lower harvesting rates, the throughput declines monotonically with increase in $P_{col}$. However, at higher harvesting rates, the trend is not monotonous and dependent on the value of $P_{col}$. Such behavior can be explained intuitively as follows: As $P_{col}$ increases (in turn, $P_{D}$ decreases), two opposite behaviors exist that affect the throughput:
1) SU attempts to transmit more number of times. However, increase in $P_{col}$ results in more number of collisions with PU and the number of failed attempts increases. Thus, the energy is wasted in failed attempts and the energy available for transmission decreases, which defines the drop in throughput. 
2) The number of samples required for sensing decreases according to \eqref{eqn:nmin}. This increases the time available for transmission, increasing the throughput. 

\begin {table}
\caption {Simulation Parameters}
\centering
    \begin{tabular}{| l | l | l | l |}
    \hline
    Notation  & Value & Notation  & Value\\ \hline
    $B$ &  200 kHz & $P_b$  & $10^{-3} $ \\ \hline
    $T_{c} $ &  1 ms & $\kappa $ & 1.9 \cite{energy_goldsmith}\\ \hline
    $ T$ &  1 ms & $ e_{s, \text{sample}} $  & $ 0.11 \times 10^{-6} $J \cite{park1}\\ \hline   
    $ P_{ckt} $  & 188 mW \cite{energy_goldsmith} & Pilot symbols/channel & 14\\ \hline
    $ f_{s} $ & 200 kHz & $ K $ & 4\\ \hline
     $\gamma$  & 0 dB & $N_{0}$ & $ 2 \times 10^{-10} $ W/Hz\\ \hline  
     \end{tabular}
     \label{tab:simparam}\vspace*{-3mm}
\end{table}

At low $P_{EH}$ (15, 60 mJ/s, Fig. \ref{collvspeh}), the effect of energy lost in collisions is more pronounced than the increased transmission duration as the energy loss cannot be compensated by the small newly harvested energy. Thus, the throughput reduces with increase in $P_{col}$. However, at high ${P}_{EH}$ (120, 180 mJ/s) and small $P_{col}$, the loss in energy is compensated. Hence, the throughput increases with $P_{col} $ due to increase in the transmission time. But, at high $P_{col}$, the effect of energy loss is more prominent than the increased transmission time even at high $P_{EH}$ reducing the throughput. Also, a trade-off exists against the variation of $P_F$. At a given harvesting rate, as $P_F$ increases, the number of minimum samples required reduces increasing the transmission duration, in turn, increasing the throughput. However, simultaneously, SU is more frequently denied access to channel even if the channel is unoccupied, reducing the throughput. 

\begin{figure}
\centering \hspace*{-15mm} 
\includegraphics[scale=0.31]{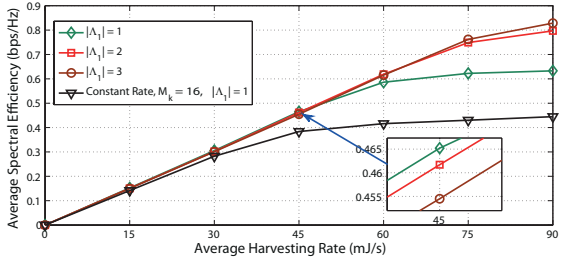}
\caption{Effect of number of channels chosen for sensing $|\Lambda_1|$ on throughput for myopic policy with $ P_{col} = P_F = 0.1 $ when the channel between the secondary pair is Rayleigh with power gain $g$ and the channel between PU and SU is AWGN, $N$ = $|\Lambda_0|$ = 5. Transition probabilities for 5 channels are  $\beta$ = [0.8 0.7 0.65 0.6 0.5] and $\alpha$ = [0.3 0.4 0.45 0.5 0.6].}
\label{diff_pfa}\vspace*{-4mm}
\end{figure}

Fig. \ref{diff_pfa} shows that at lower harvesting rates, the SU throughput when a single channel is sensed is higher than that of when upto 3 channels are sensed; while at higher harvesting rates, the throughput of the latter is higher. This is because at lower harvesting rates, much of the harvested energy and the time in a slot are spent on estimating and sensing multiple channels reducing the energy and the time available for transmission, in turn, reducing the throughput. At higher harvesting rates, enough energy is available to compensate the loss in transmission duration and the expenditure of energy in estimation and sensing giving higher throughput for multiple sensed channels as the probability of finding an available channel to access is higher compared to the case of a single sensed channel. Also, adapting to fading conditions gives better average spectral efficiency than that of the constant rate transmission.

\begin{figure}[t]
\centering \hspace*{-15mm} 
\includegraphics[scale=0.32]{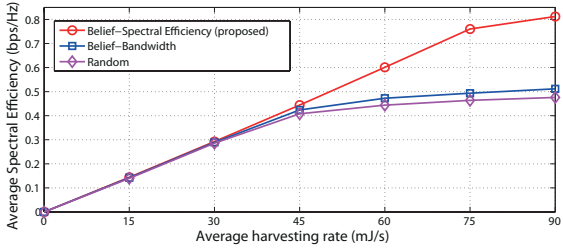}
\caption{Comparison of myopic policies with different channel selection criteria. $ P_{col}$ = $P_{F} = 0.1$, $N$ = $\vert \Lambda_{0}\vert = 6 $, $\vert \Lambda_{1}\vert = 3$. Each channel has transition probabilities $\beta = 0.7$ and $\alpha = 0.3$.}
\label{comp_crit}\vspace*{-5mm}
\end{figure}
Fig. \ref{comp_crit} shows that the proposed channel selection criterion based on belief and energy-constrained spectral efficiency is more suitable for an energy harvesting CR with fading channels compared to existing criteria like belief-bandwidth based channel selection \cite{park} and random channel selection. In belief-bandwidth criterion, the channel selection is purely based on the belief about PU occupancy and bandwidths of channels. However, as the proposed criterion exploits channel conditions and the probabilistic energy availability to make sensing decision, gain in throughput is obtained by choosing channel(s) with better gain(s) and using the energy efficiently.

\begin{figure}[t]
\centering \hspace*{-10mm} 
\includegraphics[scale=0.32]{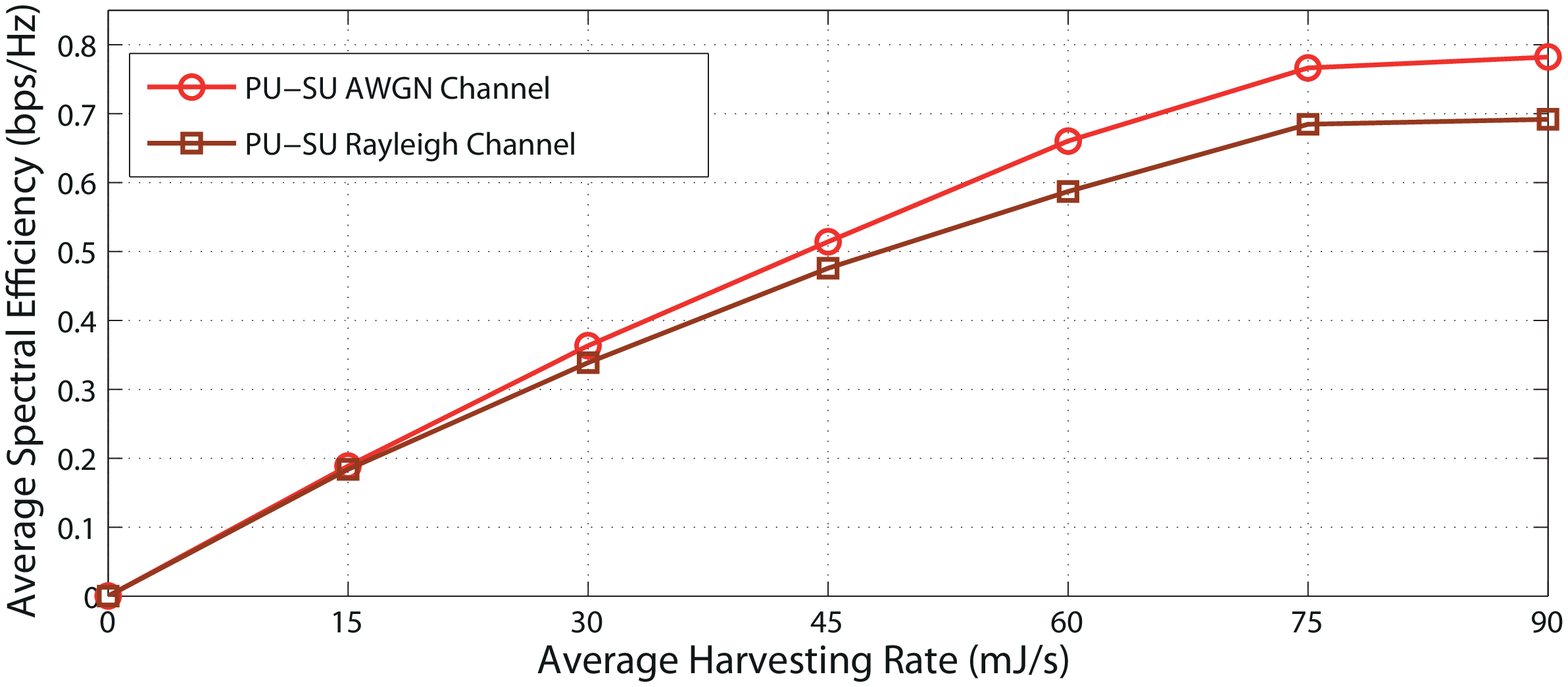}
\caption{Comparison of optimal policies for different channel coefficient distributions between PU and SU with $P_{col}$ = $P_{F} = 0.1 $. The channel between the secondary pair is Rayleigh with power gain $g$, $N$ = $\vert \Lambda_{0}\vert = 4 $, $\vert \Lambda_{1}\vert = 1$. Transition probabilities for 4 channels are $\beta$ = [0.8 0.7 0.65 0.6], $\alpha$ = [0.3 0.4 0.45 0.5].}
\label{awgnrayl}\vspace*{-5mm}
\end{figure}

When the channel between PU and SU is Rayleigh, the minimum sensing duration $T_{s,\min}$ given by \eqref{eqn:tmin} and $e_{s} $ varies in each slot with respect to $\sqrt{h}$ for the given $P_{D}$ and $P_{F}$. In fact, $T_{s,\min}$ may exceed the total time slot duration $T$ for deep fade channel conditions. In that case, no action is taken and the throughput for the corresponding slot is zero. We assume $\mathbb{E}\left[\vert h \vert \right] = 1$. As shown in Fig. \ref{awgnrayl}, fading on the sensing channel results in reduction in SU throughput as the throughput remains zero at some slots for Rayleigh channel due to longer sensing duration and more energy is consumed due to deep fade compared to AWGN channel.\vspace*{-2mm}

\bibliographystyle{ieeetr}
\bibliography{paper}

\end{document}